\documentclass[prl,aps,twocolumn]{revtex4}

\usepackage{psfig,epsf}

\begin{document}

\title{Autocorrelation exponent of conserved spin systems in the
scaling regime following a critical quench}
\author{Cl\'ement Sire}

\affiliation{
Laboratoire de Physique Th\'eorique (UMR 5152 du CNRS), Universit\'e
Paul Sabatier, 118, route de Narbonne, 31062 Toulouse Cedex 4, France\\
E-mail: {\it clement.sire@irsamc.ups-tlse.fr } }

\begin{abstract}
We study the autocorrelation function of a conserved spin system
following a quench at the critical temperature. Defining the
correlation length $L(t)\sim t^{1/z}$, we find that for times $t'$
and $t$ satisfying $L(t')\ll L(t)\ll L(t')^\phi$ well inside the
scaling regime, the spin autocorrelation function behaves like
$\langle s(t)s(t')\rangle\sim
L(t')^{-(d-2+\eta)}\left[{L(t')}/{L(t)}\right]^{\lambda^\prime_c}$.
For the $O(n)$ model in the $n\to\infty$ limit, we show that
$\lambda^\prime_c=d+2$ and $\phi=z/2$. We give a heuristic
argument suggesting that this result is in fact valid for any
dimension $d$ and spin vector dimension $n$. We present numerical
simulations for the conserved Ising model in $d=1$ and $d=2$,
which are fully consistent with the present theory.
\end{abstract}
\maketitle

The quench of a ferromagnetic spin system \cite{bray}, from high
temperature ($T_0>T_c$) to low temperature (typically $T=0$ or
$T=T_c$) is characterized by the growth of a correlation length
scale (or domain length scale when domains can be identified),
$L(t)\sim t^{1/z}$. In the non conserved case, $z$ depends on the
final temperature of the quench ($z=2$ for $T<T_c$, while $z$ is
the dynamical critical exponent for $T=T_c$ \cite{hh}). If the
order parameter $s({\bf x},t)$ (possibly a vector) is locally
conserved, $z=3$ (scalar) or $z=4$ (vector) for a quench below
$T_c$ \cite{bray}, while $z=4-\eta$ \cite{hh} for a quench at
$T_c$. Another interesting and fundamental quantity is the spin
autocorrelation $A(t,0)=\langle s({\bf x},t)s({\bf
x},0)\rangle\sim L(t)^{-\lambda}$
\cite{lambdac,lambdabibth,lambdabibexp}. For non conserved
dynamics, whatever the temperature of the quench, $\lambda$ is non
trivial (except in $d=1$ \cite{bray}) and only approximate
theories are available for $T=0$ \cite{bray,lambdabibth}, while
for $T=T_c$ \cite{lambdac}, the $\varepsilon$-expansion of
$\lambda_c$ can be calculated. In the case of conserved dynamics,
it is now well established that $\lambda_c=\lambda=d$ for quenches
at and below $T_c$ \cite{satya1,satya2,huse}. Hence, for fixed
$t'$ and $t\to+\infty$,
$A(t,t')\sim L(t)^{-d}$.
However, for $t'$ and $t>t'$ both in the scaling regime (in a
sense to be defined later), several authors have observed
numerically \cite{lambdapbib1,lambdapbib2,godreche} and
experimentally \cite{explamb} a faster power law decay of the
autocorrelation. More precisely, in the case of a quench of an
Ising system at $T_c$ (critical quench), the authors of
\cite{godreche} obtained numerically
\begin{equation}
A(t,t')\sim
L(t')^{-(d-2+\eta)}\left[\frac{L(t')}{L(t)}\right]^{\lambda'_c},
\label{lambdap}
\end{equation}
in $d=1$ (where formally $\eta=1$ and $T_c=0$) and $d=2$. They
respectively found $\lambda'_c\approx 2.5$ in $d=1$ and
$\lambda'_c\approx 3.5$ in $d=2$. They also suggested a general
scaling relation
\begin{equation}
A(t,t')\sim L(t)^{-d}C\left[\frac{L(t)}{L(t')^\phi}\right],
\label{scalam}
\end{equation}
where $C(x)$ goes to a non zero constant for $x\to +\infty$,
$C(x)\sim x^{-(\lambda'_c-d)}$ for $x\to 0$, and
\begin{equation}
\phi=1+\frac{2-\eta}{\lambda'_c-d}.\label{phigen}
\end{equation}
As noticed in \cite{godreche}, this scaling implies the existence
of a new relevant length scale $L(t')^\phi$,
which is the crossover length between the two observed
regimes. Its physical meaning has yet to be elucidated.

In the present Letter, we address the problem of the actual
analytical derivation of $\lambda'_c$ in the case of the $O(n)$
model in the limit of infinite $n$. Within this model, the
diffusive nature of this new length scale can be understood, and
one finds $\lambda'_c=d+2$ and $\phi=2$. By generalizing the
interpretation of this diffusive crossover length scale to any
$O(n)$ spin system, we conjecture that the result $\lambda'_c=d+2$
holds and that $\phi=2-\eta/2=z/2$.

We first examine the exactly solvable $O(n)$ model in the limit
$n\to\infty$ and for dimensions $d>2$. This model is known to be
pathological for a quench at zero temperature, displaying
multiscaling \cite{coniglio}, whereas normal scaling should be
restored at finite $n$ \cite{bray,bray1}. However, after a quench
at $T_c$, the structure factor obeys standard scaling even for
$n\to\infty$ \cite{satya2}.
In the standard Cahn-Hilliard equation describing the evolution of
the magnetization field ${\bf s}({\bf x},t)$, ${\bf s}^2({\bf
x},t)/n$ can be replaced by its average in the limit $n\to\infty$.
Thus, any spin component
satisfies
\begin{equation}
\frac{\partial s}{\partial t}=-\Delta\left[\Delta s+k_0^2s-\langle
s^2\rangle s\right]+\eta,\label{ch}
\end{equation}
where $k_0^2$ is a constant, $\eta({\bf x},t)$ is a conserved
delta-correlated noise satisfying $\langle\eta({\bf k},t)\eta({\bf
k'},t')\rangle=2T_ck^2\delta({\bf k}+{\bf k'})\delta(t-t')$, and
$\langle s^2\rangle$ has to be computed self-consistently.
Although the derivation of the structure factor has already
appeared in the literature \cite{satya2}, we briefly repeat it as
it furnishes a useful basis for our final derivation.

Eq.~(\ref{ch}) can be solved in Fourier space, leading to
\begin{equation}
s({\bf k},t)=\left[s({\bf k},0)+\int_0^t{\rm
e}^{q(k,\tau)}\eta({\bf k},\tau)\,d\tau\right]{\rm e}^{-q(k,t)},
\label{s}
\end{equation}
where
\begin{equation}
q(k,t)=k^4t-k^2\int_0^t[k_0^2-\langle s^2({\bf
x},\tau)\rangle]\,d\tau.
\end{equation}
Assuming an uncorrelated initial condition such that $\langle
s(-{\bf k},0)s({\bf k},0)\rangle=s_0^2$, we then find the
structure factor $S(k,t)=\langle s(-{\bf k},t)s({\bf k},t)\rangle$
\begin{equation}
S(k,t)=\left[s_0^2+2T_ck^2\int_0^t{\rm
e}^{2q(k,\tau)}\,d\tau\right]{\rm e}^{-2q(k,t)}.
\end{equation}
We now express the self-consistent condition $\langle s^2({\bf
x},t)\rangle=\int^\Lambda S(k,t)\,\frac{d^d{\bf k}}{(2\pi)^d}$,
where $\Lambda$ is the inverse of a lattice cut-off. $T_c$ is such
that $S(k,t\to\infty)\sim k^{-2+\eta}$, where $\eta$ is the usual
critical exponent controlling the decay of the static correlation
function ($\eta=0$ for $n\to\infty$). This leads to
\begin{equation}
T_c\int^\Lambda k^{-2}\,\frac{d^d{\bf
k}}{(2\pi)^d}=k_0^2.\label{tc}
\end{equation}
Finally, if the above condition is satisfied, we find that
$q(k,t)$ obeys a scaling relation for large $t$
\begin{equation}
q(k,t)=q(kt^{1/4})=k^4t-c_dk^2t^{1/2},
\end{equation}
where $c_d$ is a universal constant determined by a simple integral
relation ($c_d=0$ for $d>4$) \cite{satya2}, and $q(u)=u^4-c_du^2$. We
thus find $L(t)=t^{1/z}$, with $z=4$, in agreement with the general
result $z=4-\eta$ \cite{hh}. We hence reproduce the general form of
the structure factor
\begin{equation}
S(k,t)=s_0^2{\rm e}^{-2q(kt^{1/z})}+t^{(2-\eta)/z}F(kt^{1/z}).
\label{scas}
\end{equation}
For the  $O(n=\infty)$ model, we have $z=4$, $\eta=0$,
and
\begin{equation}
F(u)=2T_cu^2\int_0^1{\rm e}^{-2u^4(1-v)+2c_du^2(1-v^{1/2})}\,dv.
\end{equation}
Note the following asymptotics for $F(u)$
\begin{eqnarray}
F(u)&\sim &2T_c u^{2}, \quad u\to 0,\label{as0}\\
F(u)&\sim &T_c u^{-2}, \quad u\to +\infty.\label{asinf}
\end{eqnarray}
In the scaling limit, the first term of the right hand side (RHS) of
Eq.~(\ref{scas}) is negligible compared to the second term. In
real space, Eq.~(\ref{scas}) illustrates the fact that
conventional (critical) scaling is obeyed
\begin{equation}
\langle{\bf s}({\bf x},t){\bf s}({\bf
0},t)\rangle=L(t)^{-(d-2+\eta)}f[x/L(t)],\label{f}
\end{equation}
where $f$ is simply the inverse Fourier transform of $F$.

We now move to the calculation of the two-time correlation
function, focusing on the case where both considered times $t'$
and $t>t'$ are in the scaling regime, a notion which will be made
more precise hereafter. Using  Eq.~(\ref{s}), and working along
the line of the derivation of $S(k,t)$, we find the following
expression for $C(k,t)=\langle s(-{\bf k},t')s({\bf k},t)\rangle$
\begin{eqnarray}
C(k,t,t')&=& A_1(k,t,t')+A_2(k,t,t'),\\
A_1(k,t,t')&=& s_0^2{\rm e}^{-q[kL(t')]-q[kL(t)]},\label{c1}\\
A_2(k,t,t')&=& L(t')^2{\rm e}^{q[kL(t')]-q[kL(t)]}F[k
L(t')].\label{c2}
\end{eqnarray}
For a fixed $t'$ and $t\to\infty$, the contribution of $A_2$
becomes negligible, as for large $t$ and hence $L(t)$, only the
contribution of small wave vector $k\sim L(t)^{-1}$ matters. Using
the result of Eq.~(\ref{as0}), we indeed find that that this term
is of order $k^2\sim L(t)^{-2}$, whereas the main contribution
$A_1$ in Eq.~(\ref{c1}) is of order $s_0^2$ which is a constant.
Contrary to what occurs for $S(k,t)$, it is now the term depending
on the initial conditions via $s_0^2$ which dominates. Hence in
this limit of fixed $t'$ and $t\to\infty$, we find
\begin{equation}
C(k,t,t')\approx C(k,t,0)= s_0^2{\rm e}^{-q[kL(t)]}=G[kL(t)],
\end{equation}
and in real space
\begin{equation}
\langle{\bf s}({\bf x},t){\bf s}({\bf 0},t')\rangle=
L(t)^{-d}g[x/L(t)],
\end{equation}
where $g$ is  the inverse Fourier transform of $G$. One recovers,
in the limit $t\gg t'$ to be made more precise later, that the
large time autocorrelation exponent is $\lambda_c=d$, which is
observed in all conserved models including thermal fluctuations
\cite{satya1,satya2}. In this limit, conventional scaling holds.
However, we will now show that the contribution of Eq.~(\ref{c2})
which has not so far been considered dominates in a well defined
time regime, and will prompt us to introduce another
autocorrelation exponent $\lambda'_c$.

For general $t'$ and $t>t'$, we now proceed to calculate the
autocorrelation for a spin on a given lattice site. Defining
$A(t,t')=\langle s({\bf x},t')s({\bf x},t)\rangle$, we finally
find $A(t,t')=A_1(t,t')+A_2(t,t')$, where
\begin{equation}
A_1(t,t')=s_0^2\int^\Lambda {\rm
e}^{-q[kL(t')]-q[kL(t)]}\,\frac{d^d{\bf k}}{(2\pi)^d}.
\end{equation}
After a change of variable and noting that the region of $k\gg
L(t)^{-1}$ barely contributes to the integral, we find
\begin{equation}
A_1(t,t')=L(t)^{-d}a_1[L(t')/L(t)].
\end{equation}
 Thus, $A_1(t,t')$ obeys conventional
scaling for any $t'$ and $t>t'$. We explicitly find
\begin{equation}
a_1(u)=s_0^2\int^\infty {\rm
e}^{-k^4(1+u^4)+c_dk^2(1+u^2)}\,\frac{d^d{\bf k}}{(2\pi)^d},
\end{equation}
where this integral is now over the entire space.
$a_1(u)$ remains bounded and of order $s_0^2$ for any value of
$u=L(t')/L(t)\leq 1$.
Keeping the notation $u=L(t')/L(t)\leq 1$, the expression for
$A_2(t,t')$ can be written in the rescaled form
\begin{eqnarray}
&&A_2(t,t')=L(t')^{-(d-2)}u^d
{\times}\label{a2}\\
&&\int^{L(t)\Lambda}{\rm
e}^{-k^4(1-u^4)+c_dk^2(1-u^2)}F(ku)\,\frac{d^d{\bf
k}}{(2\pi)^d}.\nonumber
\end{eqnarray}

Let us analyze the different asymptotics for $A_2(t,t')$. First of
all, for large $t=t'$ ($u=1$),  the integral is dominated by the
region of large $k$'s. Using Eq.~(\ref{tc}), we find the expected
result $A_2(t,t)\approx k_0^2$, which is the equilibrium value of
$\langle s^2\rangle$. Note that if $t-t'\ll 1$, we obtain
$A_2(t,t')=A_2(t,t)-K_d(t-t')+...$, where $K_d$ is a computable
constant. We now assume that $1\ll L(t)-L(t')\ll L(t')$, which
ensure that $u$ is very close to 1. In this regime, we find that
\begin{eqnarray}
A_2(t,t')&\sim &J_d
L(t)^{-(d-2)}\left[1-\frac{L(t')}{L(t)}\right]^{-(d-2)/4}, \\
&\sim & J'_d\,(t-t')^{-(d-2)/4},
\end{eqnarray}
where $J_d$ and $J'_d$ can be written exactly as simple integrals.
Finally, and this constitutes the central result of this Letter,
we consider the regime $1\ll L(t')\ll L(t)$. In this case, $u\ll
1$, and the integral of Eq.~(\ref{a2}) is dominated by the region
of $k$ of order unity, so that the small argument asymptotics can
be taken for $F(ku)$ in Eq.~(\ref{a2}). We find
\begin{eqnarray}
A_2(t,t')&\sim &\kappa_d\,
L(t')^{-(d-2)}\left[\frac{L(t')}{L(t)}\right]^{d+2},\label{ln} \\
\kappa_d&= &2T_c\int^\infty k^2{\rm
e}^{-k^4+c_dk^2}\,\frac{d^d{\bf k}}{(2\pi)^d}.
\end{eqnarray}
Eq.~(\ref{ln}) takes exactly the expected form of
Eq.~(\ref{lambdap}), with
\begin{equation}
\lambda'_c=d+2.
\end{equation}
Hence, we find that $A_2(t,t')$ {\it prevails over} $A_1(t,t')$
for $L(t')\leq L(t)\ll L_0(t')$, with $L_0(t')\sim L(t')^\phi$ and
$\phi=2$. For $L(t')\ll L(t)\ll L_0(t')$, the autocorrelation
$A(t,t')\approx A_2(t,t')$ is then given by Eq.~(\ref{ln}).
Moreover, Eq.~(\ref{a2}) shows that instead of Eq.~(\ref{scalam}),
the correct scaling is rather
\begin{equation}
A(t,t')= A(t,0)+L(t')^{-(d-2+\eta)}D[L(t)/L(t')],\label{scageneral}
\end{equation}
with $D(1/u)\sim u^{\lambda'_c}$ for $u\ll 1$. Both scaling are
equivalent only for $u\ll 1$. We now present an heuristic argument
based on dimensional analysis which suggests that the result
$\lambda'_c=d+2$ may be of general validity for conserved spin
systems. Indeed, the occurrence of a new length scale bigger than
$L(t)$ could have been inferred from the small $k$ behavior of
$S(k,t)$. In the $n\to\infty$ limit and for $k\to 0$,
Eq.~(\ref{scas}) leads to
\begin{equation}
S(k,t)\approx s_0^2+2T_ck^2 L(t)^4+...\label{smallk}
\end{equation}
A natural momentum scale $k_0(t)\sim L_0(t)^{-1}$ arises by matching
the two terms of the RHS of Eq.~(\ref{smallk}),
which leads to $\phi=2$ and hence $\lambda'_c=d+2$.

\begin{figure}[ht]
\psfig{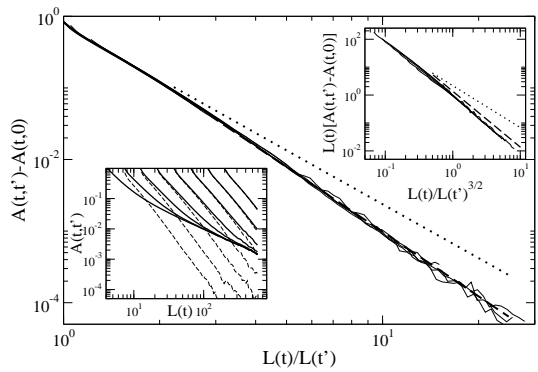} \caption[]{Illustrating the
result of Eq.~(\ref{scageneral}), we plot
$A(t,t_k)-A(t,0)=D[L(t)/L(t_k)]$, for
$L(t_k)\approx L(0)r^k$, with $r=1.75$ and $k=1,...,7$ (40000 samples
of length $N=5000$). Although the initial slope is smaller
($\lambda'_c\approx 2.5$ \cite{godreche}; dotted line), the asymptotic
exponent is very close to $\lambda'_c=3$ (dashed line fit). The bottom
inset shows the original data for $A(t,t_k)$ and $A(t,t_k)-A(t,0)$
(dashed lines).  The top inset shows $L(t)[A(t,t_k)-A(t,0)]$ as a
function of $L(t)/L(t_k)^{\phi}$ (with $\phi=3/2$). Lines of slope
$\lambda'_c-1$ are shown for $\lambda'_c=3$ (dashed line) and
$\lambda'_c\approx 2.5$ (dotted line).}
\end{figure}
In the general case, for short-range correlated initial
conditions, we expect the following general form to hold
\begin{equation}
S(k,t)=F_1[k L(t)]+L(t)^{2-\eta} F_2[k L(t)],
\end{equation}
with $F_1(0)=s_0^2$ being a non zero constant (equal to the
variance of the initial total magnetization normalized by the
volume), while the scaling contribution should vanish for $k=0$,
implying $F_2(0)=0$. Imposing $F_2(p)\sim p^\gamma$, $\gamma$ is
necessarily an even integer. If $\gamma$ were not integer, the
correlation function scaling function $f$ defined in Eq.~(\ref{f})
would have a power law decay for large distance, which is
unphysical as such correlations cannot develop in a finite time
starting from short-range ones. $\gamma$ cannot be an odd integer
as space isotropy guarantees that $f$ should be an even function.
Contrary to the case of a quench at $T=0$, for which convincing
theoretical arguments for $d\geq 2$ \cite{k4} and experiments
\cite{expk4} show that $F_2(p)\sim p^4$,
there is no reason to expect the same for critical quenches.
Generically, we expect $F_2(p)\sim p^2$ as found for the $d=1$
conserved Ising model \cite{satya1,satya2}, and in the present
Letter for the $O(n)$ model for $n\to\infty$. Finally, the small
$k$ behavior of the structure factor should be of the form
\begin{equation}
S(k,t)\approx s_0^2+C_0k^2 L(t)^{4-\eta}+...,\label{smallkgen}
\end{equation}
where $C_0>0$ is a constant. Assuming that the length scale obtained
by matching both terms of the RHS of Eq.~(\ref{smallkgen}) is the
same as the crossover length between the two observed regimes for
the autocorrelation, and using the general result of
Eq.~(\ref{phigen}), we obtain
\begin{equation}
\phi=2-\eta/2=1+\frac{2-\eta}{\lambda'_c-d},
\end{equation}
which implies $\lambda'_c=d+2$. This result also extends to $d=1$
after formally taking $\eta=1$, leading to $\lambda'_c=3$ and
$\phi=3/2$. Note that the crossover scale can also be written
$L_0(t)\sim t^{\phi/z}\sim t^{1/2}$, which is the diffusion scale
associated to thermal noise. At least in $d=1$, this scale can be
related to the equilibrium diffusion of tagged spins observed in
\cite{luck}.
\begin{figure}[ht]
\psfig{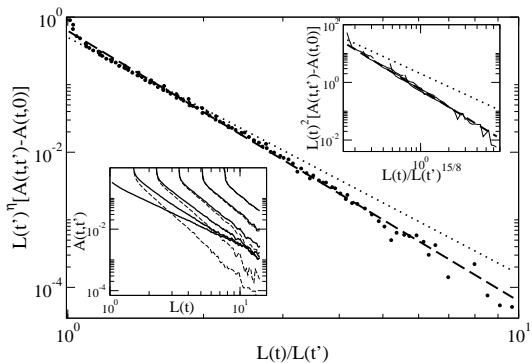} \caption[]{In the bottom inset,
we plot $A(t,t_k)$ and $A(t,t_k)-A(t,0)$ (dashed lines) as a function
of $L(t)$, for $L(t_k)\approx r^k$, with $r=1.5$ and $k=1,...,5$ (16
samples of size $N=500{\times} 500$, $L(0)=1/\sqrt{2}$). The main plot
shows $L(t_k)^\eta[A(t,t_k)-A(t,0)]=D[L(t)/L(t_k)]$. Although the
initial slope is consistent with $\lambda'_c\approx 3.5$
\cite{godreche} (dotted line), the effective exponent certainly
increases and the asymptotic slope is more compatible with
$\lambda'_c=4$ (dashed line fit).  The top inset shows
$L(t)^2[A(t,t_k)-A(t,0)]$ as a function of $L(t)/L(t_k)^{\phi}$ (with
$\phi=15/8$). Although not as clean as in $d=1$, the scaling plot is
better described by the line corresponding to $\lambda'_c=4$ (dashed
line) rather than $\lambda'_c\approx 3.5$ (dotted line).}
\end{figure}
We now present simulations of the Ising model Kawasaki dynamics in
$d=1$ and $d=2$ after a quench at $T_c$. In the $d=1$ case, we use
the accelerated algorithm introduced in \cite{satya2}, which is
faster than that used in \cite{godreche} (but does not permit to
compute simply the response function as was needed in
\cite{godreche}). By fitting $A(t,t')$ in the scaling regime, the
authors of \cite{godreche} found $\lambda'_c\approx 2.5$ lower
than our prediction $\lambda'_c=3$.  However, for the moderately
large numerically accessible times, the contribution of
$A_1(t,t')\approx A(t,0)$ is significant. When plotting
$A(t,t')-A(t,0)$ as a function of $L(t)$, one actually finds
$\lambda'_c\approx 3$ instead of $\lambda'_c\approx 2.5$.
Result of simulations for the $d=2$ Ising model evolving with
Kawasaki dynamics at $T_c$ are shown on Fig.~2. Considering the
very slow growth of $L(t)\sim t^{4/15}$, it is difficult to obtain
data spanning more than one decade in $L(t)$. Hence, the regime of
interest $1\ll L(t')\ll L(t)$ cannot be reached and the separation
of scales properly achieved. Still, subtracting $A(t,0)$ from
$A(t,t')$ leads to $\lambda'_c\approx 4$, significantly greater
than the value $\lambda'_c\approx 3.5$ found in \cite{godreche}.

In conclusion, in view of the exact result for the $O(n=\infty)$
model, a general argument for any $n$ and $d$, and convincing
simulations in $d=1$ (and consistent in $d=2$), we have strongly
suggested that $\lambda'_c=d+2$ and $\phi=z/2$ generally holds. We
also find that the scaling form of Eq.~(\ref{scageneral}) is more
appropriate than Eq.~(\ref{scalam}).  The compelling
generalization of our heuristic argument to a quench at $T<T_c$
(in $d\geq 2$, and admitting $F_2(p)\sim p^4$) leads to
$A(t,t')\sim [L(t')/L(t)]^{\lambda'}$ for $L(t')\ll L(t)\ll
L(t')^\phi$, with $\lambda'=d+4$ and $\phi=1+d/4$. In $d=2$, the
prediction $\lambda'=6$ is significantly larger than the numerical
result $\lambda'\approx 4$ \cite{lambdapbib1}.  However, the fit
in \cite{lambdapbib1} was performed in the short scaling regime
over less than a decade in $L(t)$, and subtracting $A(t,0)$ before
performing the fit could lead to a significantly higher value for
$\lambda'$, as noted in the two examples treated in this Letter.

\end{document}